\documentclass[11pt]{article}
\usepackage{amsmath,amsfonts,amsthm,bm,slashed,bbm,mathrsfs}
\usepackage{graphicx,color}
\newcommand{\nn}{\nonumber}

\newcommand{\ben}{\begin{enumerate}}\newcommand{\een}{\end{enumerate}}
\newcommand{\Ref}[1]{(\ref{#1})}

\newcommand{\be}{\begin{equation}}\newcommand{\ee}{\end{equation}}
\newcommand{\bea}{\begin{eqnarray}}\newcommand{\eea}{\end{eqnarray}}

\newcommand{\la}{\lambda}\newcommand{\La}{\Lambda}
\newcommand{\ga}{\gamma}
\newcommand{\Ga}{{E_p}}

\newcommand{\om}{\omega}
\renewcommand{\sp}{\slashed p}\newcommand{\sk}{\slashed k}

\newcommand{\p}{\mathbf{p}}\newcommand{\pp}{\underline{p}}
\renewcommand{\k}{\mathbf{k}}\newcommand{\kk}{\underline{k}}

\newcommand{\eq}[2]{\begin{align}\label{#1}#2\end{align}}

\begin{document}

\title{Photon dispersion relations in $A_0$-background}

\author{M. Bordag\thanks{bordag@uni-leipzig.de}\\
{\small Universit{\"a}t Leipzig, Institute for Theoretical Physics,  Germany},\\
V.~Skalozub\thanks{skalozubv@daad-alumni.de}\\
{\small Oles Honchar Dnipro National University, Dnipro, Ukraine}}
\date{Sept. 21, 2018 
}
\maketitle\thispagestyle{empty}

\section*{Abstract}
We calculate the photon  dispersion relations generated by the quark loop   in a quark-gluon plasma  with   the color  $A_0$ background condensate $A_0^c = A_0^{c3} + A_0^{c8}$ = const. It is found
that both transversal and longitudinal modes are exited.   They have a gap at low momenta and  are stable in high temperature approximation. The background fields act as imaginary chemical potentials and decrease the photon frequencies compared to the
 case of zero background. The comparison with QED plasma with chemical potential is discussed.

\section{\label{T1}Introduction}
 Searching for  new state of matter, quark-gluon plasma (QGP), is in
the center of  modern high energy physics. According to  present
day knowledge, it consists of quarks and gluons  (or corresponding
quasi-particles) deliberated from hadrons. As numerous
calculations (analytic, numeric, lattice simulations, combined
lattice and analytic)  showed this state is not a free particle
gas. The QGP background  has  a rather complicated structure,
which is  formed from a gluon field condensate, so-called $A_0$
condensate, and spontaneously created chromo-magnetic fields
\cite{skal1708.02792}. The appearance of these classical fields lowers
the free energy of the plasma.  This background, in particular,
results in $Z(3)$ center symmetry breaking at high temperature
\cite{anis84-10-423}. The  deconfinement phase transition order
parameter  is Polyakov's loop \cite{mcle81-98-195}.  It is the
integral over temporal component of gluon field,
\be \label{PL} P(\vec{x}) = T \exp\bigl[i g \int dx_4 A_0
(\vec{x}, x_4) \bigr], \ee
which  equals zero in confining and nonzero in deconfining phases.
Various aspects of the $A_0$ condensation are widely discussed in
the literature (see review paper \cite{bori95-43-301}). The best developed part concerns the thermodynamical properties of QGP in
this environment. Different models for the effective potential
\cite{pisa00-62-111501},  \cite{dumi02-525-95}, \cite{elze87-35-748},
\cite{sasa1204.4330}, \cite{meis02-66-105006} were proposed. In particular, which
is important for us here, the temperature interval where the presence of the $A_0$ condensate
 is dominant, has been estimated to be $T  \sim T_d - 2.5 T_d$,
where $T_d$ is deconfinement temperature \cite{meis04-585-149}. At
higher temperatures, the gluon quasiparticle contributions are
more significant. The spectra of gluons in the  $A_0$ background have
also been calculated \cite{kala96-11-1825}. It was discovered, in
particular,  that either transversal or longitudinal modes are
excited. The later ones resemble plasmons in QED plasma.

In what follows, we  restrict ourselves to this interval of temperatures. Here, if necessary, the magnetic fields could be accounted for in perturbation theory.

Another important aspect of the $A_0$ condensate is related to the propagation and scattering processes of different particles which result in new type phenomena. These may serve as signals of the QGP creation. Scattering on the $A_0$ condensate as a classical external field may result in $C$-parity violating processes \cite{atre1111.3027} and others.

On the other hand,  Polyakov's  loop is an extended object which is not a solution to local  field equations. So, the profile of the $A_0$ configuration is not specified and could be any, in general. In such a situation, below we consider the configuration with a constant potential, $A_0$ = const.  It is a  solution of the local field equations and can be derived from an effective two-loop potential \cite{anis84-10-423}, or by using other field theoretic methods (see \cite{bori95-43-301}). Its relation to Polyakov's loop is obvious. In fact, it is a good approximation for studying    different  processes in QGP.

In the present paper, we calculate and investigate the dispersion relations for photon in QGP in a $A_0$ = const background. These calculations are similar to that in \cite{kala96-11-1825}  and  fill a gap in the  literature. They  are also important for the phenomenology of QGP since for the detection of the plasma state we have to look for related photon modes which could exist in both the plasma and the vacuum. The quarks carry both electric and color charges and  the quark loops modify the photon spectra in dependence on the color  background.

The paper is organized as follows. In next section we give  necessary information on quark interactions in the $A_0$ background and the  notations used. In sect. 3 the one-loop photon polarization tensor (PT) is calculated. Sections 4 and 5 are devoted to photon dispersion relations in the high temperature approximation  and for intermediate temperatures, correspondingly. Conclusions and discussion are given in the last section.
\\Throughout the paper we use units with $\hbar=c=k_{\rm B}=1$.

\section{ Lagrangian and basic formulas}
We consider  the color background field described by the potential $
B_\mu^c = \delta_{\mu  4} (  A_0^{c3} + A_0^{c8}) = const$.
Euclidean space-time is used. The Lagrangian describing
interactions of quarks with electromagnetic and gluon fields reads
\be \label{IntV} L = - \bar{\psi} \gamma_\mu V_\mu \psi  -  m
\bar{\psi}  \psi.\ee
The matrix condensed notations are introduced.
The two expressions describing electromagnetic  and strong
interactions are joined  in \Ref{IntV}:
\begin{eqnarray}  (V_\mu)^{f f'}_{i j} = \left[\begin{array}{cc}  (a_\mu)^{f f'}_{i j} & 0 \\
 0 & (Q_\mu)^{f f'}_{i j} \end{array} \right].
\end{eqnarray}

 The matrix of electromagnetic interactions  has the form
\be \label{Mem} (a_\mu)^{f f'}_{i j} = e ~ r_f A_\mu \delta^{f f'}
\delta^{i j}, \ee
and the one describing strong interactions is
\be \label{Mg}  (Q_\mu)^{f f'}_{i j} = g  Q^{i j}_\mu \delta^{f
f'} , \ee
where the color matrix is $Q^{i j}_\mu =
\frac{(\lambda^a)^{ij}}{2} Q^a_\mu$ and $\la^a$ - Gell-Mann
matrixes, $A_\mu$ and $Q_\mu$ denotes  potential of
electromagnetic and gluon field, correspondingly.  In \Ref{Mem}, \Ref{Mg} we  marked color by Latins
i, j,... = 1, 2, 3.
%
%
The mass matrix is  also  flavor dependent: $ m = m_f \delta_{f
f'}$.
The quark wave function can be taken in the form $ ([\psi)^f_i]^T
= ( \psi_i^u, \psi_i^d, \psi_i^s, ...| \psi^f_{i = 1}, . \psi^f_{i
= 2}, .. ., \psi^f_{i = 3})$. In the expressions \Ref{Mem} and
\Ref{Mg}, in one matrix    either the  flavor or the color
variables are joined. We note that a specific flavor has the
corresponding electric charge. In units of proton charge we have
$e_f = | e |  r_f$ and the number $r_f$ corresponds to this
flavor. For example, for $u$-quark $r_u = 2/3$, for $d$ quark $r_d
= - 1/3$,  for s-quark  $r_s = - 1/3$, etc. The electromagnetic
field feels the flavor parameter $e_r$ of a quark whereas the
gluon feels color (strong) charge $g$. In terms of these  objects all the calculations can be
carried out.  Below we restrict ourselves to three light quarks,
only.

As we see from Eq.\Ref{Mem}, electromagnetic interaction is
diagonal  with respect to color and flavor variables. To account
for the background color fields we take into consideration the
structure of the $\la^a$ matrixes. In QCD the $ A_0$ was
calculated already by analytic methods of field theory in either
gluon or quark sector (see \cite{bori95-43-301}). In these cases, two
background fields could be generated - $A_0^3$ and $A_0^8$,
correspondingly  to the diagonal generators $T^3 =
\frac{\la^3}{2}$,  $T^8 = \frac{\la^8}{2}$,
\begin{eqnarray} \label{gelllanm}
\la^3 = \left[\begin{array}{ccc}
1& 0&0\\
0&- 1&0\\
0&0&0
\end{array}\right], ~~\la^8 = \frac{1}{\sqrt{3}} \left[\begin{array}{ccc}
1& 0&0\\
0& 1&0\\
0&0& - 2
\end{array}\right].
\end{eqnarray}

The background fields $B^c_\mu = \delta_{\mu 0}(\delta^{c3}A^3_0 +
\delta^{c8}A^8_0)$ belong to the gauge group center:
\be \label{A0} A_0  = \frac{2 \pi n}{g \beta},~~  n \to Z_3: \ee
$\beta = 1/T$, T is temperature.

In  \cite{skal94-9-4747} - \cite{skal93-57-324} it was obtained that in
two-loop approximation the  condensed fields are:
\be \label{fieldB} g A_0^3 = \frac{g^2 T}{4 \pi} (3 - \xi),~~
A_0^8 = 0.\ee
$\xi$ is gauge fixing parameter.

We have to expect that $A_0^8 \not = 0$ in general. It can happen
in higher  orders in loop expansion.  It is also important that
the value of the created field does not depend  on the quark mass.

For our problem with photon PT, the actual values of the
background fields are  not very essential and  so we  consider two
scenarios: 1) The situation happening in the case of \Ref{fieldB};
2) the  case  $B_0^3, B_0^8 \not = 0$.

 Since the latter case is more general, we continue for this one.
Let us  take into account the light quark flavors: $ f = u (m_u,
e_u = 2/3 e),~  d (m_d, e_d = - 1/3 e),~ s (m_s, e_s = - 1/3 e)$.
$e $ is proton electric charge.
 Accounting for the matrix structure of the generators $T^3,   T^8$ \Ref{gelllanm},
 we have to put the background fields  for quark flavors:
  $B_0^u = (+ A_0^3 + A_0^8/\sqrt{3})/2, 2/3  e,   m_u$  for u-quark;  $B_0^d = (- A_0^3 + A_0^8/\sqrt{3})/2, - 1/3  e , m_d$
  for d-quark  and  $B_0^s =  - A_0^8/\sqrt{3},  e_s = - 1/3  e,  m_s$~ for s-quark.
  These have to   be multiplied by the  coupling g in the covariant derivatives.

Now, we turn to calculation of one-loop photon PT at these
background at high temperature $T > T_d$.  Standard imaginary time
formalism is used.
 In this formulation,
the fermion propagator at $A_0$ background has a fourth momentum
component $p_4=2\pi T l-A_0$ with half-integer $l$. In what
follows the  notations will be used:
 $p_\mu=(p_4,\p)_\mu$ and $\pp$ is $p_\mu$ written
without index,  $\p$ is spatial part and $p=|\p|$. The same
conventions are used for $k$.

\section{\label{T2}Photon polarization tensor}
We start from the Schwinger-Dyson equation for the photon
propagator (PT),
\eq{1}{{ \mathscr{D}^{}}^{-1}_{\mu\nu} &=
{\mathscr{D}^{(0)}}^{-1}_{\mu\nu}-\Pi_{\mu\nu}(\kk) }
where ${\mathscr{D}^{(0)}}^{-1}_{\mu\nu}$ is  inverse free field propagator.
The photon PT accounting  for a one flavor is given by
\eq{2}{\Pi^{\rm G}(\kk)_{\mu\nu}&=
    -e^2T\sum_{l=-\infty}^{\infty} \int\frac{d^3p}{(2\pi)^3}
    {\rm Tr}\,\ga_\mu  \frac{1}{\sp-m} \ga_\nu  \frac{1}{\sp-\sk-m},
\\\nn   &=e^2T\sum_{l=-\infty}^{\infty} \int\frac{d^3p}{(2\pi)^3}
    \frac{Z_{\mu\nu}}{(p^2+m^2)((p-k)^2+m^2)},
}
where the minus sign is from the Fermion loop and $e = e_f$ is
the electric charge. In the following we drop the index '$f$' and the sum over '$f$'. In the second line we defined
\eq{3}{Z_{\la\la'}&=-{\rm Tr}\ga_\mu(\sp+m)\ga_\nu(\sp-\sk+m) }
for the polynomial  in the numerator. Carrying out the trace (it
is Euclidean) we get
\eq{4}{\frac14 Z_{\mu\nu}&=
    \delta_{\la\la'}(\pp(\pp-\kk)+m^2)-p_{\la}(p-k)_{\la'}-(p-k)_{\la}p_{\la'}.
}
Below we need the combinations
\eq{5}{\frac14 Z_{44}&=-p_4(p-k)_4+\p(\p-\k)+m^2,
\\[4pt]\nn   \frac14 Z_{\mu\mu}&=2\pp(\pp-\kk)+4m^2.
}
We divide the expression \Ref{1} for the polarization tensor
into vacuum and temperature parts,
\eq{6}{\Pi_{\mu\nu}(\kk)  &= \Pi^{\rm vac.}_{\mu\nu}(\kk)+\Delta_T
\Pi_{\mu\nu}(\kk). } The vacuum part has the simple tensor
structure
\eq{7}{\Pi^{\rm vac.}_{\mu\nu}(\kk) &=
\left(\delta_{\mu\nu}{\kk^2}- {k_\mu k_\nu}\right)\Pi(\kk^2) } and
the representation in terms of a parametric integration of
$\Pi(\kk^2)$ is
\eq{8}{ \Pi(\kk^2) &= -\frac{e^2}{8\pi^2}
        \int_0^1 dx \,x(1-x)\ln\left(1-x(1-x)\frac{\kk^2}{m^2}\right).
} In the temperature dependent part we use the following formula.
Let
\eq{9}{ \Pi_{\mu\nu}(\kk) &= T e^2 \sum_{l=-\infty}^{\infty}
\int\frac{d^3p}{(2\pi)^3}
    \frac{Z(p_4,\p)_{\mu\nu}}{(p^2+m^2)((p-k)^2+m^2)}
}
be the analytic expression for a one-loop graph with two lines
and $Z(p_4,\p)_{\mu\nu}$ be the polynomial in the numerator,
where we explicitly indicated its dependence on the integration
momenta. Then
\eq{10}{
    \Delta_T\Pi_{\mu\nu}(\kk)&=e^2
    \int\frac{d^3p}{(2\pi)^3} \frac{1}{2\Ga}
  \left\{
    n_s\left[\frac{Z_{\mu\nu}(i\Ga,\p)+Z_{\mu\nu}(-i\Ga-k_4,-\p-\k)}{N}+{\rm c.c.}\right]
    \right.\\\nn&\left.~~~~~~~~~~~~~~~~~~~~~~~~~~~~~~~~~~~~~~~~~
    +
    n_a\left[\frac{Z_{\mu\nu}(i\Ga,\p)-Z_{\mu\nu}(-i\Ga-k_4,-\p-\k)}{N}-{\rm c.c.}\right]
    \right\},
}
with
\eq{10a}{\Ga=\sqrt{p^2+m^2}, \ \  N =  \kk^2+2ik_4\Ga+2\p\k, }
and
\eq{11}{
     n_s=\frac12(n_{A_0}+n_{-A_0}), \ \ n_a=\frac12(n_{A_0}-n_{-A_0}), \ \  n_{A_0}=\frac{- 1}{\exp(\beta(\Ga+iA_0))+ 1}
}
is the combination with the Boltzmann factor for fermions. The
notation $ +  c.c. $ prescribes to add the complex
conjugated  of the term given in the bracket.  The background field $A_0$
appears only in the Boltzmann factors. In the following, the
antisymmetric combinations, i.e., the second term in the square
bracket in \Ref{10}, will disappear after taking the angular integrations. We
mention that these formulas are formally the same as for the QED
in a dense medium with the substitution $\mu\to iA_0$ for the
chemical potential. In case of $A_0$-condensate, however, $
\Delta_T\Pi_{\mu\nu}(\kk)$ is a periodic function of $A_0$ with
period $2\pi T$.

Applying \Ref{10} to our expressions \Ref{5} we note the
substitutions
\eq{12}{\frac14 Z_{44}&\to N-4ik_4\Ga+4\Ga^2-\kk^2,
\\[4pt]\nn   \frac14 Z_{\mu\mu}&\to 2(N-\kk^2+2m^2).
}
and get from \Ref{2} and \Ref{5}
\eq{13}{  \Delta_T\Pi_{44}&=
        4e^2\int\frac{d^3p}{(2\pi)^3}\frac{n_s}{2\Ga} \left(1-\frac{\kk^2 -4\Ga^2+4ik_4\Ga}{N}+{\rm c.c.}\right),
\\[4pt]\nn   \Delta_T\Pi_{\mu\mu}&=
        4e^2\int\frac{d^3p}{(2\pi)^3}\frac{n_s}{2\Ga} 2\left(1-\frac{\kk^2-2m^2}{N}+{\rm c.c.}\right).
}
In these formulas the angular integration can be carried out
using known formulas (see, e.g., \cite{kala84-32-525})
\eq{14}{  \int_{0}^\pi d\theta \sin(\theta)\frac{1}{N}
    =\frac{\ln(a)-\ln(b)}{4pk}
}
with the angle $\theta$ between the vectors $\p$ and $\k$ such
that $\p\k=pk\cos(\theta)$ holds and the notations
\eq{15}{\ln a
&=\frac{(\kk^2+2pk)^2+4k_4^2\Ga^2}{(\kk^2-2pk)^2+4k_4^2\Ga^2}, \ \
\ln b = \ln\frac{\kk^4-4(pk-ik_4\Ga)^2}{\kk^4-4(pk+ik_4\Ga)^2} }
were introduced.
We mention that the above expressions   were initially defined for real
$k_4$ and allow for direct continuation to real frequencies,
$k_4\to -i \om$.

Using with \Ref{10a} the relation
\eq{16}{ 2\p\k=N-\kk^2-2ik_4\Ga, } after angular integration by means of \Ref{14} the
expressions \Ref{13} turn into
\eq{14a}{       \Delta_T \Pi_{44}(k) &=
        \frac{2}{\pi^2}e^2\int_0^\infty \frac{ dp\,p^2}{\Ga}\,n_s M_{44},
\nn\\       \Delta_T \Pi_{\mu\mu}(k) &=
        \frac{2}{\pi^2}e^2\int_0^\infty  \frac{ dp\,p^2}{\Ga}\,n_s M_{\mu\mu},
\nn \\
 \Delta_T A(k) &=
        \frac{2}{\pi^2}e^2\int_0^\infty  \frac{ dp\,p^2}{\Ga}\,n_s M_A,
} with
\eq{15a}{ M_{44} &= 1-\frac{(\kk^2
-4\Ga^2)\ln(a)-4ik_4\Ga\ln(b)}{8pk}, \nn\\  M_{\mu\mu} &=
1-\frac{(\kk^2-2m^2)\ln(a)}{8pk}, \nn\\   M_{A} &=
    1-\frac{k_4^2}{k^2}
    +\frac{(k_4^4-k^4-4p^2\kk^2-4\Ga^2k_4^2)\ln(a)
    -4ik_4\Ga\kk^2\ln(b)}{8pk^3},
} where we also displayed the expressions for the combination $A=
\frac12\Pi_{\mu\mu}-\frac{\kk^2}{2\k^2}\Pi_{44}$, which will
appear below for the transversal mode.

Further we have to consider the tensor structures. We use the following notations,
\eq{16a}{ A_{\mu\nu}&=\delta_{\mu\nu}-\frac{k_\mu k_\nu}{\kk^2}
                                                    -h_\mu h_\nu,
\ \    B_{\mu\nu} = h_\mu h_\nu,
        \ \  h_\mu=\frac{k_4}{k\kk}k_\mu-\frac{\kk}{k}u_\mu,\ \
    u_\mu=\delta_{\mu 4}
}
for the two independent tensor structures which we have at finite temperature. The polarization tensor has the decomposition
\eq{17}{ \Pi_{\mu\nu}=A_{\mu\nu} \Pi_t+ B_{\mu\nu} \Pi_{l}
}
with
\eq{18}{ \Pi_l &=\frac{\kk^2}{k^2}\Pi_{44},
       \ \  \Pi_t \equiv A= \frac12\Pi_{\mu\mu}-\frac{\kk^2}{2k^2}\Pi_{44}.
}
With these notations and the free inverse photon propagator,
\eq{19}{{\mathscr{D}^{(0)}}^{-1}_{\mu\nu} =
(\delta_{\mu\nu}\kk^2-k_\mu k_\nu),
}
one has the (also well known) formula
\eq{20}{\mathscr{D}_{\mu\nu}   &=
    \frac{1}{\kk^2-A}A_{\mu\nu}
    +\frac{k^2/\kk^2}{k^2-\Pi_{44}}B_{\mu\nu}
}
whose poles determine the spectrum. The excitations which go
with $A_{\mu\nu}$  are transversal (there are two of them with
equal spectrum) and that going with $B_{\mu\nu}$ are longitudinal.

Splitting the vacuum part of the polarization tensor according to \Ref{17}, we get from \Ref{7} and \Ref{16a}
\eq{21}{  \Pi^{\rm vac.}_{\mu\nu}(\kk) &=
        ( A_{\mu\nu}+ B_{\mu\nu})\kk^2 \Pi(\kk^2)
}
and together with \Ref{18} we arrive at
\eq{22}{A &= \kk^2 \Pi(\kk^2)+\Delta_T A, \ \
    \Pi_{44} = k^2\Pi(\kk^2)+\Delta_T \Pi_{44}.
}
Finally, from \Ref{20} the dispersion equations
\eq{23}{ \kk &= \kk^2\Pi(\kk^2)+\Delta_T A,\ \ \
        k^2 = k^2\Pi(\kk^2)+\Delta_T \Pi_{44}
}
follow.

\section{\label{T3}Spectra at high temperature}
In this section we consider the spectrum in leading order for
$T\to\infty$.  In this case expressions significantly simplify. First of
all, since the temperature dependent part of the polarization
tensor is of order $T^2$ (see below), the vacuum part is
subleading. In the temperature dependent part, the leading order
comes from the expansions of $M_{44}$ and $M_A$, \Ref{15}, for
$\p\to\infty$,
\eq{24}{ M_{44}&=M_{44}^{\rm lead.} +\dots\,,\ \ \mbox{with} \ M_{44}^{\rm lead.} =  2+\frac{\om}{k}\ln\frac{\om-k}{\om+k},
\nn\\   M_A&= M_A^{\rm lead.}+\dots\,,\ \ \mbox{with} \ M_A^{\rm lead.}= \frac{\om^2}{k^2}-\frac12\left(1-\frac{\om^2}{k^2}\right)\frac{\om}{k}
               \ln\frac{\om-k}{\om+k},
}
which can be also represented using the function (eq. (2.2.3) in \cite{kala84-32-525})
\eq{25}{ F(x) &= -\frac{x}{2}
    \left(\ln\left|\frac{x-1}{x+1}\right|+i\pi\Theta(1-x)\right)
}
as
\eq{26}{  M_{44}^{\rm lead.}  &= 2\left(1-F\left(\frac{\om}{k}\right) \right),
\nn\\   M_A^{\rm lead.} &= \left(\frac{\om}{k}\right)^2+\left(1-\left(\frac{\om}{k}\right)^2\right)
F\left(\frac{\om}{k}\right).
}
The last formulas give at once the analytic continuation to $\om>k$.

After this expansion, the integrals over $p$ in \Ref{14} can be carried out and we arrive at
\eq{27}{   \Delta_T \Pi_{44}(k) &=\La^2     M_{44}^{\rm lead.} +\dots\,,
\nn \\       \Delta_T A(k) &=  \La^2 M_A^{\rm lead.}+\dots\,,
}
where we defined
\eq{28}{  \Lambda^2 &= -\frac{2e^2}{\pi^2}\int_0^\infty \frac{dp\,p^2}{\Ga}\,n_s
=  \frac{2e^2}{\pi^2}\left(\frac{\pi^2T^2}{12}
-\frac{m^2+A_0^2}{4}\right)  =
e^2\left(\frac{T^2}{6}-\frac{m^2+A_0^2}{2\pi^2 }\right),
}
which is the remaining integration over $p$ and involved the Boltzmann factors defined in \Ref{11}. In the right side is the expansion for large $T$ assuming $m$ and $A_0$ are of order $T$.
In case  $m=0$ and arbitrary $A_0$ we get for large $T$
\eq{28a}{\La^2 &=\frac{T^2}{\pi^2}
\left(-{\rm Li}_2\left(-e^{iA_0/T}\right)
-{\rm Li}_2\left(-e^{-iA_0/T}\right) \right),
}
which is a periodic function of $A_0$ and takes also negative values.

\begin{figure}[h]
    \includegraphics[width={0.6\textwidth}]{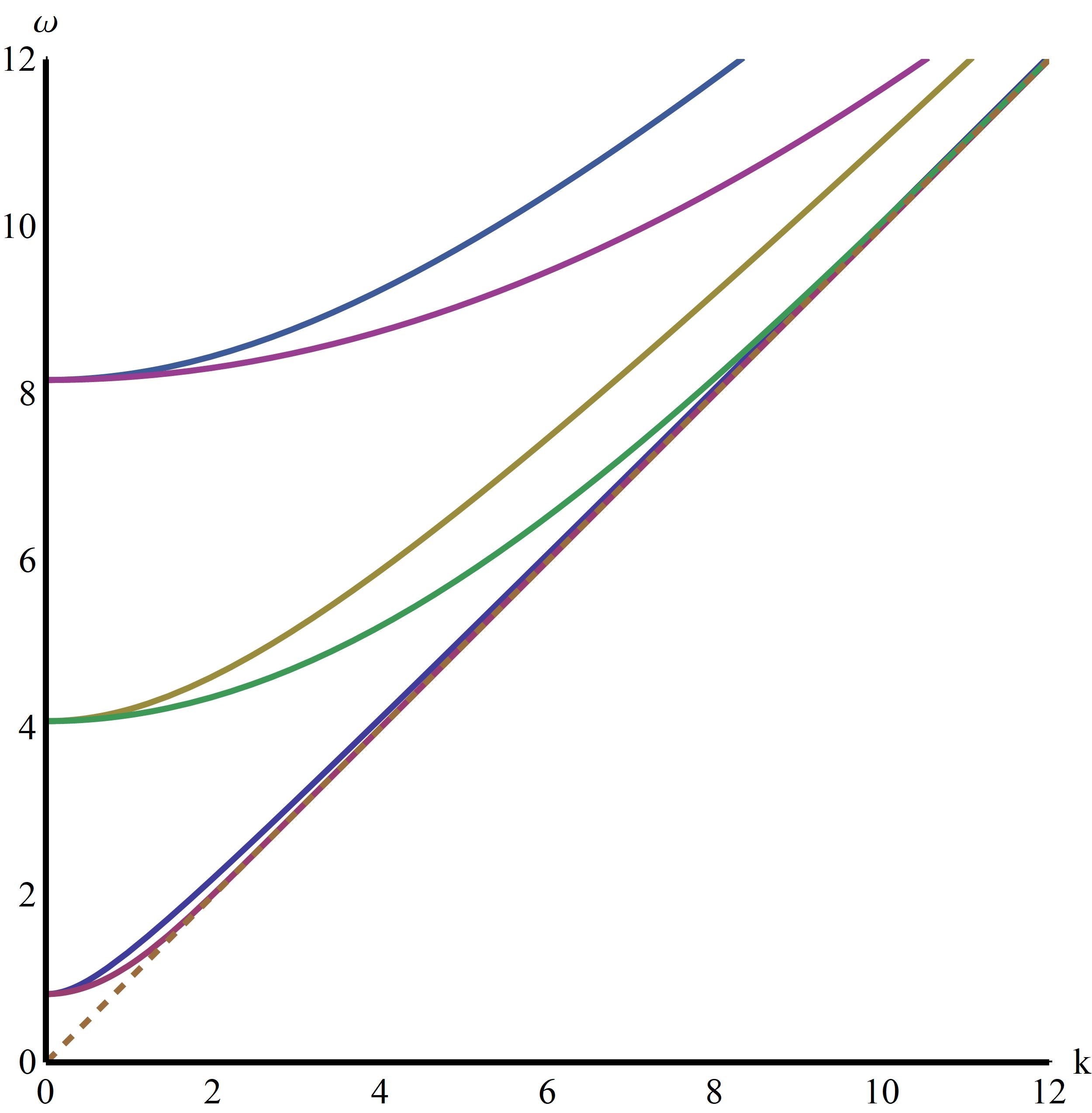}
    \caption{Solutions  $\om_ T (k) $ (upper curve in each pair) and $\om_ L (k) $ (lower curve in each pair) of the dispersion relations \Ref{23} for high temperature as function of $k$ for $\Lambda=10,\ 5,\ 1$ (from top to bottom). All solutions approach $\om=k$ for $k\to\infty$ (dashed line) and start in $\om_{T,L}(0)=\sqrt{\frac23}\La $ (see Eq. \Ref{32}).
    }\label{fig1}
\end{figure}

\begin{figure}[h]
    \includegraphics[width={0.6\textwidth}]{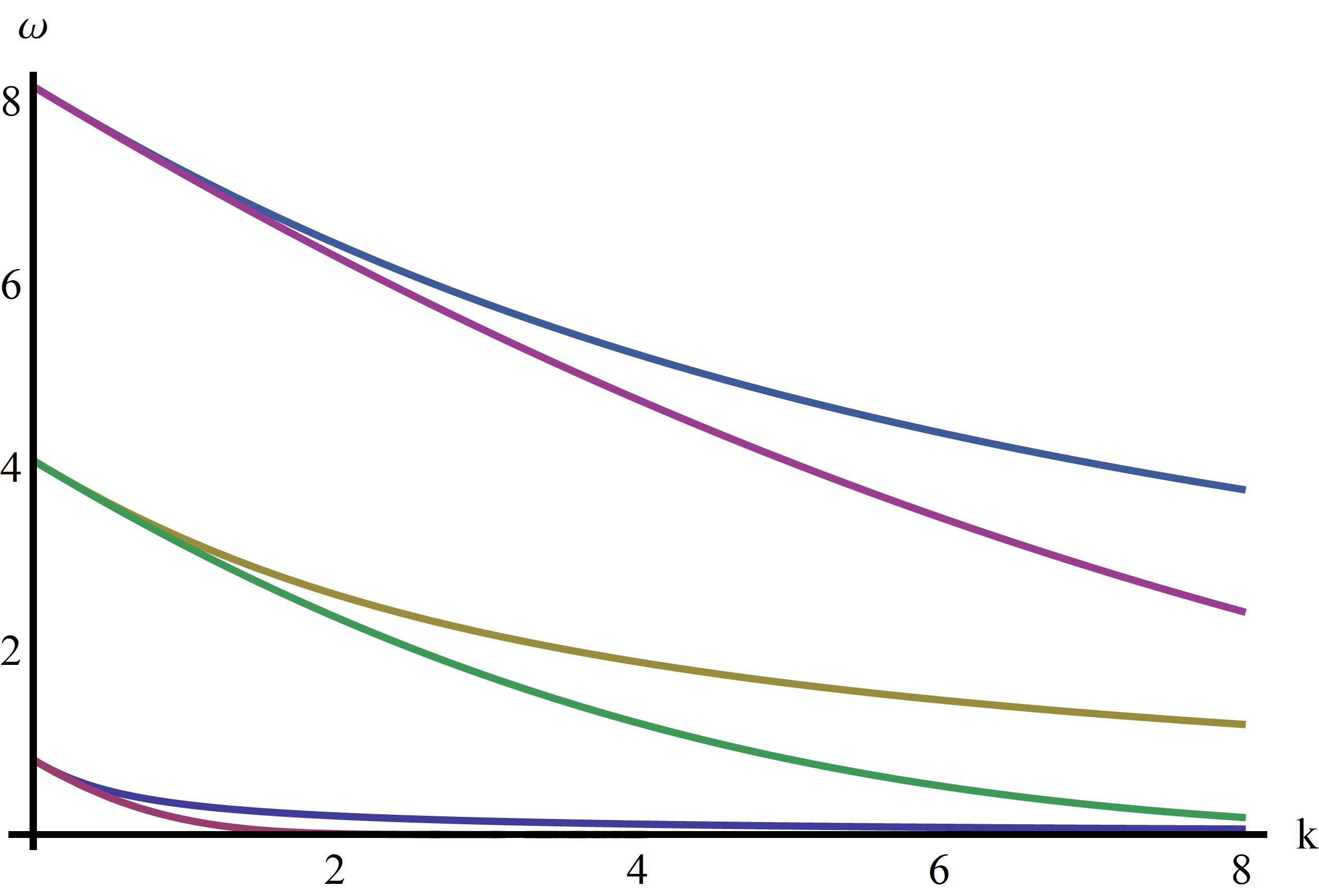}
    \caption{Deviation of the solutions of the dispersion relations \Ref{23} from momentum $k$,  $\om_ T (k)-k $ (upper curve in each pair) and $\om_ L (k) -k$ (lower curve in each pair)  as function of $k$ for $\Lambda=10,\ 5,\ 1$ (from top to bottom).
    }\label{fig2}
\end{figure}

Turning in the dispersion equations \Ref{23} to real frequency we get for $T\to\infty$  the equations
\eq{29}{\om^2 &
=k^2+\Lambda^2\left(\frac{\om^2}{k^2}-\frac12\left(1-\frac{\om^2}{k^2}\right)\frac{\om}{k}
\ln\frac{\om-k}{\om+k}\right),
\nn\\
k^2& =-\Lambda^2\left(2+\frac{\om^2}{k^2}\ln\frac{\om-k}{\om+k}\right).
}
These equations have a scale invariance.  We define
\eq{30}{\om=\Lambda \tilde{\om}, \ \ \ {k}=\Lambda \tilde{k}
}
and the equations  turn into
\eq{31}{\tilde{\om}^2 &= \tilde{k}^2-\left(\frac{\tilde{\om}^2}{\tilde{k}^2}-\frac12\left(1-\frac{\tilde{\om}^2}{\tilde{k}^2}\right)
\frac{\tilde{\om}}{\tilde{k}}
\ln\frac{\tilde{\om}-\tilde{k}}{\tilde{\om}+\tilde{k}} \right)
\nn\\
\tilde{k}^2&{
=-\left(2+\frac{\tilde{\om}^2}{\tilde{k}^2}
\ln\frac{\tilde{\om}-\tilde{k}}{\tilde{\om}+\tilde{k}}\right)
}.}
We denote the solutions of these two equations by $\tilde{\om}_{T}(\tilde{k})$ and $\tilde{\om}_{L}(\tilde{k})$.
Their expansions for small $k$ can be obtained by iteration and read
\eq{32}{ \tilde{\om}_{T} (\tilde{k})&= \sqrt{\frac23}\left(1+\frac{9}{10}\tilde{k}^2+\dots\right),
\nn\\
\tilde{\om}_{L} (\tilde{k})&= \sqrt{\frac23}\left(1+\frac{3}{10}\tilde{k}^2+\dots\right).
}
The behavior for large $\tilde{k}$ is
\eq{33}{ \tilde{\om}_{T} (\tilde{k})&= \tilde{k}\left(1+\frac{1}{\tilde{k}^2}+\dots\right),
\nn\\
\tilde{\om}_{L} (\tilde{k})&= \tilde{k}\left(1+2e^{-\tilde{k}-2}+\dots\right).
}
The inequality $\tilde{k}<\tilde{\om}_{L}<\tilde{\om}_{T}$ holds. It must be mentioned that the expansions for $k\to\infty$ are valid for the solutions of the  equations \Ref{29}, but not for the equations \Ref{23} since in this section we assume $T$ to be the largest quantity. We mention that up to notations these formulas are in agreement with those in \cite{kala98-58-310}, which were derived for QED in a dense medium.

Restoring the  dependence on $\Lambda$ using \Ref{28}  and denoting the solutions of \Ref{27} by $\om_{{T},{L}}(k)$ we arrive at
\eq{34}{ \om_{T}(k) &= \Lambda \tilde{\om}_{T}\left(\frac{k}{\Lambda}\right),
\ \ \om_{L}(k) = \Lambda \tilde{\om}_{L}\left(\frac{k}{\Lambda}\right).
}
These functions are shown in Fig. \ref{fig1} for several values of $\Lambda$, \Ref{28}. As can be seen, the condensate $A_0$ lowers the photon frequency. The gap in both spectra is $\frac23\La$. In the given high-$T$ approximation both spectra are stable (an imaginary part appears only in next-to-leading order). Fig. \ref{fig2} shows the deviations of the solutions from the line $\om=k$.

\section{\label{T4}Imaginary part at high temperature}
As seen from the solution shown in Fig. \ref{fig1} and from the equations \Ref{29}, there is no imaginary part to order $T^2$. Thus we consider the next order, i.e., the order $T$, where an imaginary part may appear. We put $m=0$ for simplifying expressions. In order to derive these we return to eqs. \Ref{14a} and consider the continuation $k_4\to -i(\om+i0)$. We represent \Ref{15a} in the form
\eq{4.1}{ M_{44} &=1+ M_{44}^a \ln(a)+M_{44}^b\ln(b),
\nn\\   M_A&=1+\frac{\om^2}{k^2}+M_A^a\ln(a)+M_A^b\ln(b),
}
where
\eq{4.2}{  M_{44}^a &= \frac{1}{8pk}\left(\om^2-k^2+4\Ga^2\right),
 \ \ \ M_{44}^b =\frac{\om\Ga}{2pk}
 \nn\\  M_{A}^a &= \frac{1}{8pk^3}\left(-\om^4-k^4-4p^2(-\om^2+k^2)+4\om^2\Ga^2\right),
 \ \ \ M_{A}^b =\frac{\om(\om^2-k^2)\Ga}{2pk^3}.
}
Further we need the imaginary parts of $\ln(a)$ and $\ln(b)$. For that we return to \Ref{15} and represent
\eq{4.3}{\ln(a) &= \ln\left(\frac{(2p-(\om-k))(2p-(-\om-k))}{(2p-(\om+k))(2p-(-\om+k))}\right),
\nn \\ \ln(b) &=
2\ln\left(\frac{\om-k}{\om+k}\right)
+\ln\left(\frac{(2p-(\om+k))(2p-(-\om-k))}{(2p-(\om-k))(2p-(-\om+k))}\right).
}
Starting point for the analytic continuation in \Ref{4.3} is eq. \Ref{14}, where for $p\to\infty$ and $\om>k$ we have $\ln(a)\simeq 0$ and $\ln(b)\simeq 2\ln\left(\frac{\om-k}{\om+k}\right)$ which is real. Starting from here, the continuation is done with $\Im(\om)>0$ and results in,\\for $\om<k$,
\eq{4.4}{\Im(\ln(a)) &= i\pi\Theta(-\om+k<2p<\om+k),\ \
        \Im(\ln(b)) =i\pi\Theta(\om+k<2p),
}
and for $\om>k$,
\eq{4.5}{\Im(\ln(a)) &= i\pi\Theta(\om-k<2p<\om+k),\ \
        \Im(\ln(b)) =-i\pi\Theta(\om-k<2p<\om+k).
}
We insert these and \Ref{4.1} into \Ref{14a} and get for the imaginary parts,
\\for $\om<k$,
\eq{4.6}{ \Im(\Pi_{44}) &=
\frac{1}{\pi}\int_{-\om+k}^{\om+k}\frac{dp\, p^2}{\Ga}n_s
M_{44}^a
+ \frac{1}{\pi}\int_{\om+k}^{\infty}\frac{dp\, p^2}{\Ga}n_s M_{44}^b,
\nn\\ \Im(\Pi_{A})&=
\frac{1}{\pi}\int_{-\om+k}^{\om+k}\frac{dp\, p^2}{\Ga}n_s M_{A}^a
+  \frac{1}{\pi}\int_{\om+k}^{\infty}\frac{dp\, p^2}{\Ga}n_s M_{44}^b,
}
and for $\om>k$,
\eq{4.7}{ \Im(\Pi_{44}) &=
\frac{1}{\pi}\int_{-\om+k}^{\om+k}\frac{dp\, p^2}{\Ga}n_s
\left(M_{44}^a- M_{44}^b\right),
\nn\\ \Im(\Pi_{A})&=
\frac{1}{\pi}\int_{-\om+k}^{\om+k}\frac{dp\, p^2}{\Ga}n_s
\left(M_{A}^a- M_{44}^b\right).
}
These expressions behave very differently for the regions with frequency above $k$ and below. First we discuss the behavior for $\om>k$, which is the region where we found the solutions shown in Fig. \ref{fig1}. To calculate the imaginary parts for high temperature we expand the function $n_s$, \Ref{11},
\eq{4.8}{ n_s=\frac{T}{\Ga^2+A_0^2}+O(1).
}
This expansion can be inserted into \Ref{4.7} since the integration region is finite. From \Ref{4.7} we get
\eq{4.9}{ \Im(\Pi_{44}) &=
\frac{T}{\pi}\int_{-\om+k}^{\om+k}\frac{dp\, p^2}{\Ga(\Ga+A_0^2)}
\left(M_{44}^a- M_{44}^b\right),
\nn\\ \Im(\Pi_{A})&=
\frac{T}{\pi}\int_{-\om+k}^{\om+k}\frac{dp\, p^2}{\Ga(\Ga+A_0^2)}
\left(M_{A}^a- M_{44}^b\right).
}
These formulas demonstrate that the imaginary part on the solutions \Ref{34} is of order $T$, i.e., is of subleading order.

As concerns the region with $\om<k$, as seen from eq. \Ref{4.6}, we have contributions where the integration over $p$ goes up to infinity. Here we cannot use the expansion \Ref{4.8}. Instead we have to do the substitution $p\to Tp$ and must expand the factors in the parenthesis in the integrand for large argument in the same manner as we did in section \ref{T3}. As a result we get contributions of order $T^2$, which is in agreement with the imaginary part which we would get in \Ref{24} for $\om>k$ (see Eq. \Ref{25}). However, this is not the region where we have the solutions of the dispersion relations and thus this region is not physical.

\section{\label{T5}Spectra at finite temperature}
In this section we discuss some topics related to the spectra at finite temperature. This means, we do not make the approximation resulting in \Ref{26}. First of all we have to consider the vacuum contribution \Ref{7}. We rewrite the first equation   \Ref{23} with $k_4=-i\om$,
\eq{35}{\om^2(1-\Pi(\kk^2)) &= k^2 (1-\Pi(\kk^2))+\Delta_T A.
}
\begin{figure}[h]
    \includegraphics[width={0.6\textwidth}]{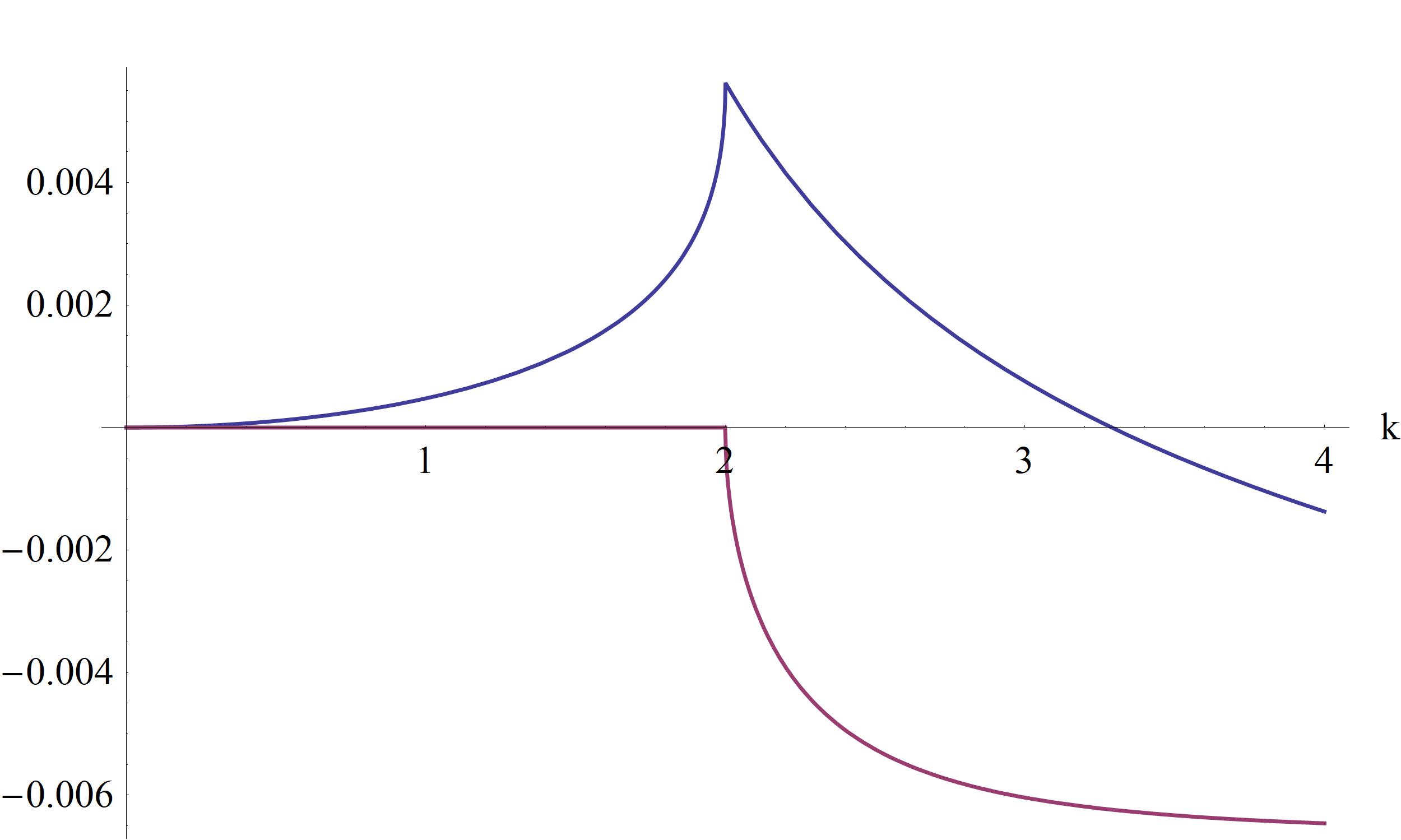}
    \caption{Real and imaginary parts of the vacuum contribution \Ref{35} to the polarization tensor for $m=1$, $e=1$.    }\label{figPI}
\end{figure}
\begin{figure}[h]
    \includegraphics[width={0.48\textwidth}]{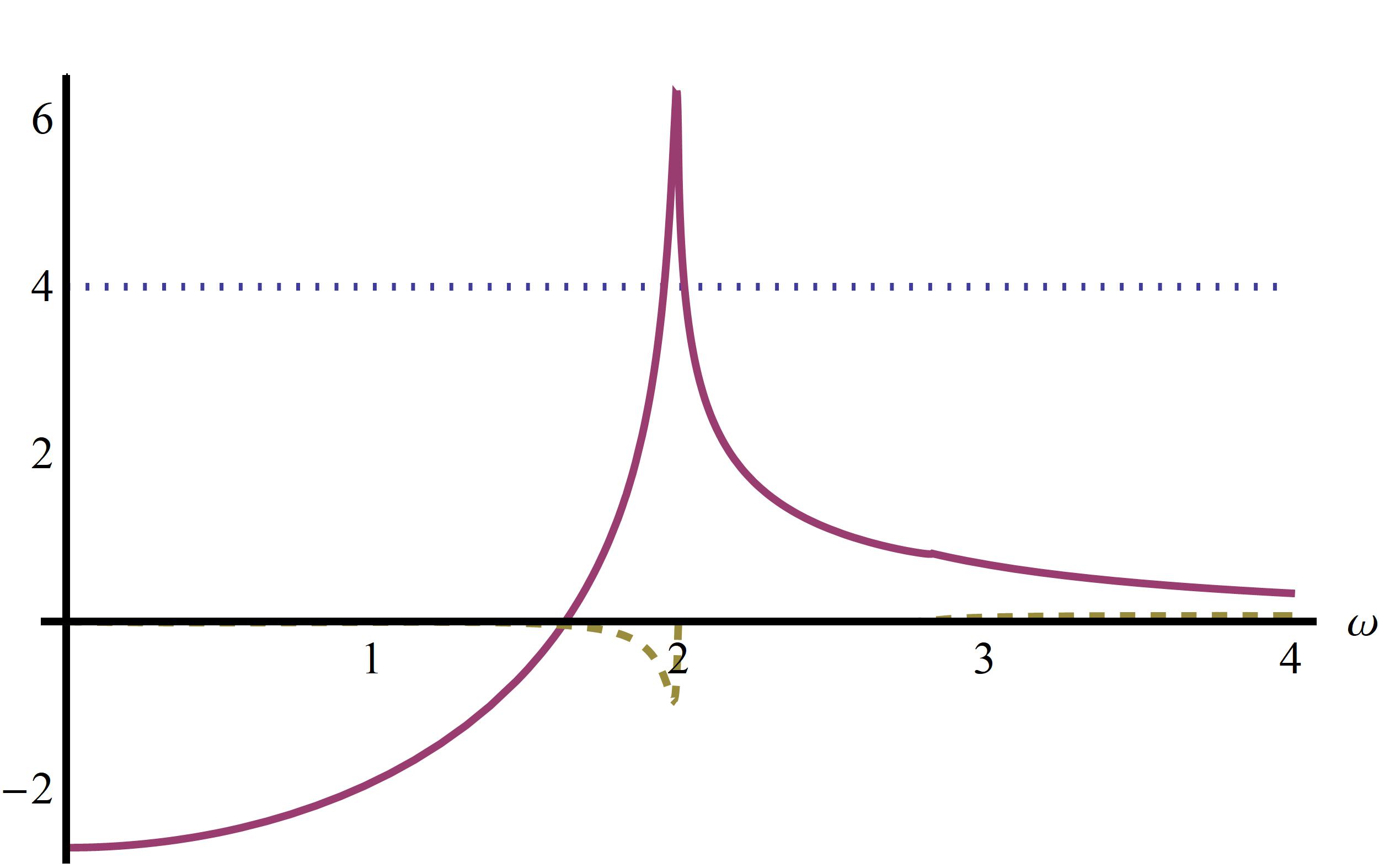}
     \includegraphics[width={0.48\textwidth}]{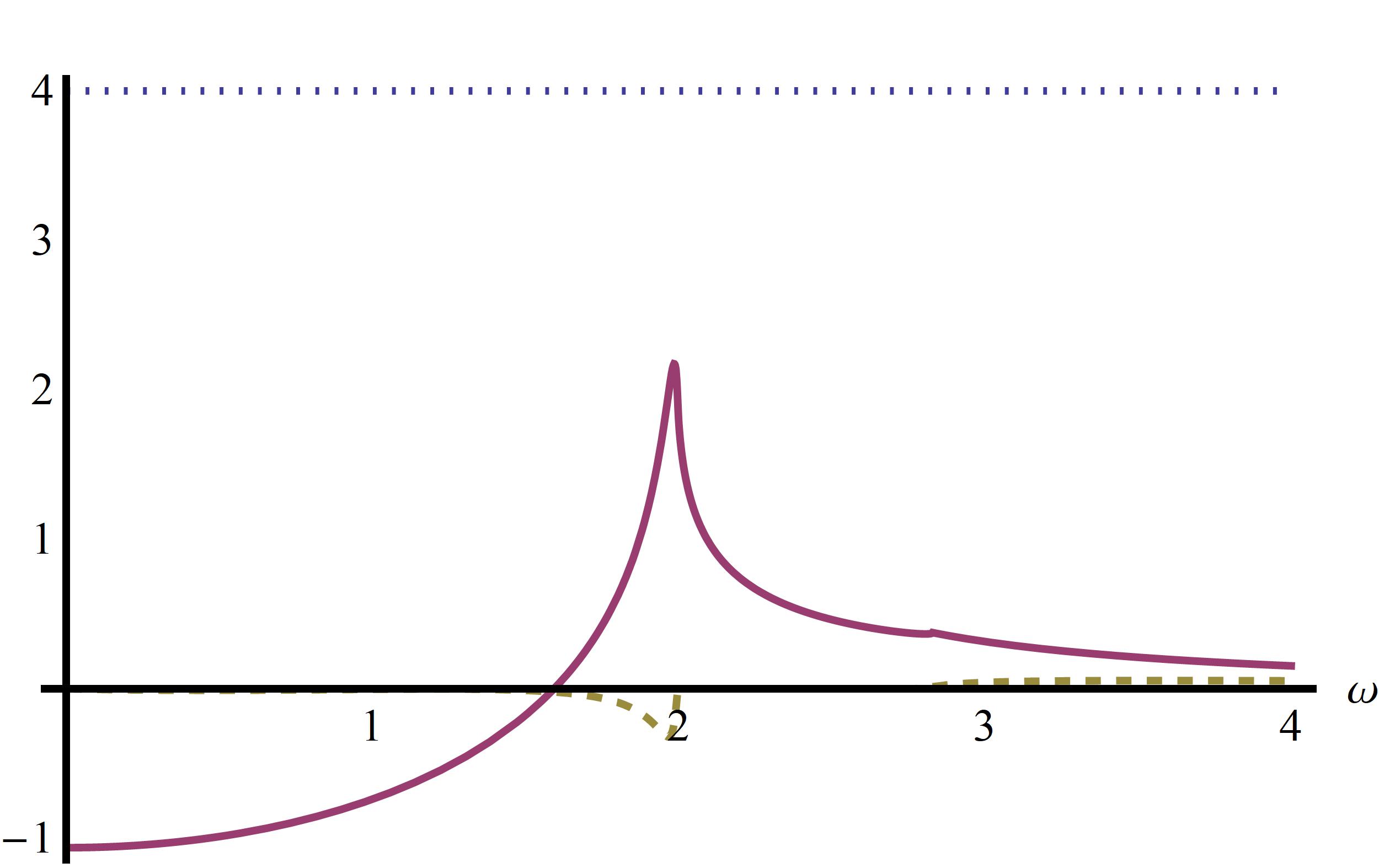}
    \caption{Left side (dotted  line) and right side (solid line for real part, dashed line for imaginary part)  of equation \Ref{35} for the longitudinal dispersion relation for $m=1$, $e=1$.  The temperature is $T=3$ in the left panel and $T=2$ in the right panel.   }\label{figLow}
\end{figure}

The function $\Pi(\kk^2)$ can be easily plotted, see Fig. \ref{figPI}. It is seen that it is a quite small quantity (unless the coupling is extremely large). For large $k$ it has the behavior $\Pi\sim-\frac{1}{24\pi^2}\ln(k)$, thus $1-\Pi(\kk^2)$ does not change sign and we can divide by this factor. In general, it is useful to remember that without temperature contributions, the only solution is $\om=k$ as it must be.  Thus our equation is
\eq{36}{ \om^2 &= k^2+\frac{\Delta_T A}{1-\Pi(\kk^2)}
}
and the solutions are similar to those in the high temperature case unless the temperature goes below $m$, where $\Delta_T A$ becomes exponentially small and we are left with the vacuum case.

It is also interesting to consider the fate of the longitudinal solution for finite $T$. From eq. \Ref{23} by the same reasons as above we have
\eq{37}{ k^2 &= \frac{\Delta_T \Pi_{44}}{1-\Pi(\kk^2)}.
}
The right side of this equation can be plotted and it has, as function of $\om$, a maximum at $\om=k$, see Fig. \ref{figLow}. The height of this maximum is given by
\eq{38}{\Delta_T \Pi_{44}(\om=k) &=
\frac{2}{\pi^2}\int_0^\infty  \frac{ dp\,p^2}{\Ga}\,n_s
\left(1-\frac{\Ga}{p}\ln\frac{\Ga+p}{\Ga-p}\right)
}
(the vacuum part does not contribute). This quantity is small for small $T$ and growing with $T$. The critical temperature is reached for
\eq{39}{k^2 &= \Delta_T \Pi_{44}(\om=k).
}
Below this temperature there is no longitudinal solution, see right panel in Fig. \ref{figLow}.

In the left panel of Fig. \ref{figLow} it is seen that there are two solutions above the critical temperature. The right one has $\om>k$ and it is that which was discussed above. The other solution, for $\om<k$, has an imaginary part and is not stable. Its existence was shortly mentioned in \cite{kala98-58-310} (after eq.(25)).

\section{Conclusions and discussion}
In two  previous sections,  we investigated in details the photon
spectra in QGP with  accounting for the presence of the background
$A_0$ fields,  which are the unavoidable  constituents of the
plasma. The condensate lowers a free energy and removes a
fictitious pole  in the  gluon spectra \cite{kala96-11-1825},
\cite{bori95-43-301}. Such a vacuum is a good approximation for
studying photon modes which have to exist in plasma and radiate
from it. The standard methods of field theory at finite
temperature were used.

As we have seen, formally the presence of the  $A_0$ background looks like
an  imaginary chemical potential $\mu_f = i g B_f$, f = u, d, s.
Hence, the photon plasma in QCD  can be presented as the set of
QED plasma constituents with different $\mu_f$. This simple
picture is qualitatively useful for either description of the
plasma properties or understanding the differences existing
between these two states of hot  matter.

First worth mentioning is that chemical  potential in QED is the
difference between the number of electrons and positrons. In QCD,
we have $\mu_f$ for quarks, only.

Next,   we have  investigated the  main sector of the center
$Z(3)$ for SU(3)  color group. The presence of the $A_0$ background breaks
this symmetry. The  other five vacua have the same energy and  can
be obtained by rotation on the angle $\pi/3$ in the color space.

It was shown  that both,  the transversal  and the longitudinal
photon modes, exist in the plasma. These spectra were
investigated in Sect. 4 in high temperature approximation and
for intermediate temperatures  in Sect. 5. It was  discovered
that the $A_0$ condensate enters the scaling $\Lambda$ factor
Eq. \Ref{28} with negative sign that lowers the photon
frequency. There exist a threshold for frequency and modes with
lower frequencies cannot propagate in the plasma.  The
$u$-quark contribution is dominant due to the electric charge
factor $e^2_f$.   This kind of  behavior  is opposite  to  QED,
where chemical potential  $\mu$ has  positive sign and the
frequency increases (compare to the zero potential case). Other
point is that there are no imaginary part in the PT in high
temperature approximation. The spectrum is stable.  The
imaginary part (and instability) appears in next-to-leading
order. This  is similar to the QED case.

In reality, $A_0$  background is not an arbitrary parameter. It has the order $A_0 \sim g T$ as typical  quantities in  temperature field theory. So, we can see that the  numerical values of the $A_0$ dependent parameters are not much changed compare to the zero condensate case. But this is important for applications because transversal photons coming out from the  plasma with the condensate  are stable objects, which could be  put in one to one correspondence with  the vacuum  photons (and wise versa). The existence of the threshold for generation of longitudinal photon modes and its dependence on the $A_0$ is also important. It gives a scale for corresponding processes in the plasma.


\begin{thebibliography}{10}

\bibitem{anis84-10-423}
R~Anishetty.
\newblock {Chemical potential for SU(N)-infrared problem}.
\newblock {\em Journal of Physics G: Nuclear Physics}, 10(4):423, 1984.

\bibitem{atre1111.3027}
Abhishek Atreya, Ajit~M. Srivastava, and Anjishnu Sarkar.
\newblock {Spontaneous CP violation in quark scattering from QCD Z(3)
  interfaces}.
\newblock {\em Phys. Rev.}, D85:014009, 2012.

\bibitem{bori95-43-301}
O.~A. Borisenko, J.~Bohácik, and V.~V. Skalozub.
\newblock {$A_0$ Condensate in QCD}.
\newblock {\em Fortschritte der Physik/Progress of Physics}, 43(4):301--348,
  1995.

\bibitem{dumi02-525-95}
Adrian Dumitru and Robert~D Pisarski.
\newblock Degrees of freedom and the deconfining phase transition.
\newblock {\em Physics Letters B}, 525(1):95 -- 100, 2002.

\bibitem{elze87-35-748}
Hans-Thomas Elze, David~E. Miller, and Krzysztof Redlich.
\newblock Gauge theories at finite temperature and chemical potential.
\newblock {\em Phys. Rev. D}, 35:748--752, Jan 1987.

\bibitem{kala84-32-525}
O.~K. Kalashnikov.
\newblock {QCD at finite temperature}.
\newblock {\em Fortsch. Phys.}, 32:525, 1984.

\bibitem{kala96-11-1825}
O.~K. Kalashnikov.
\newblock {Selfenergy peculiarities of the hot gauge theory after symmetry
  breaking}.
\newblock {\em Mod. Phys. Lett.}, A11:1825--1834, 1996.

\bibitem{kala98-58-310}
O.~K. Kalashnikov.
\newblock {Photon and Electron Spectra in Hot and Dense QED}.
\newblock {\em Physica Scripta}, 58:310, 1998.
\newblock arXiv:hep-ph/9802427.

\bibitem{mcle81-98-195}
Larry~D. McLerran and Benjamin Svetitsky.
\newblock {A Monte Carlo study of SU(2) Yang-Mills theory at finite
  temperature}.
\newblock {\em Physics Letters B}, 98(3):195 -- 198, 1981.

\bibitem{meis02-66-105006}
Peter~N. Meisinger and Michael~C. Ogilvie.
\newblock {The Finite temperature SU(2) Savvidy model with a nontrivial
  Polyakov loop}.
\newblock {\em Phys.~Rev.~D}, 66:105006, 2002.

\bibitem{meis04-585-149}
P.N. Meisinger, M.C. Ogilvie, and T.R. Miller.
\newblock {Gluon quasiparticles and the Polyakov loop}.
\newblock {\em Phys.~Lett.~B}, {585}({1-2}):{149--154}, {2004}.

\bibitem{pisa00-62-111501}
Robert~D. Pisarski.
\newblock {Quark gluon plasma as a condensate of SU(3) Wilson lines}.
\newblock {\em Phys.~Rev.~D}, 62:111501, 2000.

\bibitem{sasa1204.4330}
Chihiro Sasaki and Krzysztof Redlich.
\newblock {An effective gluon potential and hybrid approach to Yang-Mills
  thermodynamics}.
\newblock {\em Phys. Rev.}, D86:014007, 2012.

\bibitem{skal1708.02792}
V.~Skalozub and P.~Minaev.
\newblock {Magnetized quark-gluon plasma at the LHC}.
\newblock 2017.
\newblock arXiv: 1708.02792.

\bibitem{skal94-9-4747}
V.~V. Skalozub.
\newblock {Gauge invariance of the gluon field condensation phenomenon in
  finite temperature QCD}.
\newblock {\em Int. J. Mod. Phys.}, A9:4747--4758, 1994.

\bibitem{skal93-57-324}
V.V. Skalozub and I.V. Chub.
\newblock 2-loop contribution of quarks to the condensate of the gluon field at
  finite temperatures.
\newblock {\em Physics of Atomic Nuclei}, 57:324, 1993.

\end{thebibliography}
\end{document}